\documentclass[a4paper]{jpconf}
\usepackage{graphicx}
\usepackage{citesort,float}
\begin{document}
\title{Study of nucleonic matter with a consistent two- and three-body
  perturbative chiral interaction}

\author{L Coraggio$^{1}$, J W Holt$^{2}$, N Itaco$^{1,3}$, R
  Machleidt$^{4}$, L E Marcucci$^{5,6}$ and F Sammarruca$^{4}$} 

\address{$^1$ Istituto Nazionale di Fisica Nucleare, Sezione di Napoli \\
Complesso Universitario di Monte  S. Angelo, Via Cintia - I-80126 Napoli,
Italy}
\address{$^2$ Department of Physics, University of Washington \\
Seattle, Washington 98195, USA}
\address{$^3$ Dipartimento di Fisica, Universit\`a
di Napoli Federico II, \\
Complesso Universitario di Monte  S. Angelo, Via Cintia - I-80126 Napoli,
Italy}
\address{$^4$ Department of Physics, University of Idaho\\
Moscow, Idaho 83844, USA}
\address{$^5$ Istituto Nazionale di Fisica Nucleare, Sezione di Pisa\\
Largo Bruno Pontecorvo 3 - I-56127 Pisa,
Italy}
\address{$^6$ Dipartimento di Fisica ``Enrico Fermi'', Universit\`a
di Pisa, \\
Largo Bruno Pontecorvo 3 - I-56127 Pisa,
Italy}

\ead{luigi.coraggio@na.infn.it}

\begin{abstract}
We calculate perturbatively the energy per nucleon in infinite
nuclear matter with a chiral N$^3$LO (next-to-next-to-next-to-leading
order) two-body potential plus a N$^2$LO three-body force (3BF). 
The 3BF low-energy constants which cannot be constrained by two-body
observables are chosen such as to reproduce the $A=3$ binding
energies and the triton Gamow-Teller matrix element. 
This enables to study the nuclear matter equation of state in a
parameter-free approach. 
\end{abstract}

\section{Introduction}
The study of infinite nuclear matter is a very interesting topic for
its connections with the properties of other physical systems. 
As a matter of fact, for instance, the equation of state (EOS) of pure
neutron matter (PNM) may be related to the features of supernova
explosions and neutron stars, while the compressibility and the
symmetry energy of symmetric nuclear matter (SNM) are linked to giant
dipole excitations and to the neutron-proton radii in atomic nuclei,
respectively.     
This is why nowadays a great effort is currently being made to theoretically
study infinite nuclear matter using high-precision nucleon-nucleon
($NN$) potentials based on chiral perturbation theory (ChPT)
\cite{EM03,EGM05,ME11}, that may provide a clear link between nuclear physics
and quantum chromodynamics (QCD) (see for instance
\cite{Sammarruca12,Holt13b,Tews13,Krueger13,Gezerlis13,Carbone13,Baardsen13,Kohno13}).

In the framework of ChPT, nuclear two- and many-body forces are
generated on an equal footing \cite{Wei92,Kol94,ME11}.
As a matter of fact, most interaction vertices that appear in the
three-nucleon forces (3NF) and in the four-nucleon forces (4NF)
are present already at the two-body level.
In this context, a major issue is the choice of the low-energy
constants (LECs) related to the interaction vertices.
The LECs corresponding to the two-nucleon-force (2NF) vertices are
fitted to two-nucleon data, and consistency requires that these values
are not changed for the same vertices appearing in the 3NF, 4NF, \ldots
As regards the other LECs, it is worth to note that the calculation of
the EOS for pure neutron matter, with chiral 3NF
up to N$^2$LO, depends only on constants that are fixed in the
two-nucleon system. This comes out since, in this case, the contact interaction,
$V_E$, and the  1$\pi$-exchange term, $V_D$, that appear in the
N$^2$LO three-body force, vanish \cite{HS10}. 
This is not the case in SNM, whose EOS calculation is influenced by
the intermediate-range 1$\pi$-exchange component $V_D$ and the
short-range contact interaction $V_E$ of the 3NF.    
Therefore, the LECs of $V_E$ and $V_D$ (known as $c_E$
and $c_D$), which are not constrained by two-body observables, affect
the calculation of the ground-state energy of SNM. Their values should
be fixed taking into account only $A \leq 3$ observables, to avoid
biases induced by additional many-body contributions.  
Since the reproduction of the observed $A=3$ binding energies is not
sufficient to fully constrain both $c_D$ and $c_E$, 
recently, the attention has been focused on the fitting of the
triton half-life, specifically the Gamow-Teller matrix element
\cite{Gazit08,Marcucci12}, as first suggested by G\aa{}rdestig and
Phillips \cite{Gardestig06}.

In the present paper we study the EOS of both SNM and PNM employing
two- and three-nucleon chiral forces with consistent LECs, aiming to
ascertain the possibility to obtain realistic nuclear matter
predictions with chiral interactions constrained by the properties of
the two- and the three-nucleon systems without any additional adjustment.   
We employ a chiral potential with a cutoff parameter $\Lambda = 414$
MeV \cite{Cor07} and calculate,
including 3NF effects, the energy per nucleon for infinite nuclear
matter at nuclear densities in the framework of many-body perturbation
theory.
It is worth pointing out that this $NN$ potential has been already employed
in perturbative many-body calculations for finite nuclei providing a successfull 
reproduction of their spectroscopic properties \cite{Cor07,Coraggio10a}.
The effects of the N$^2$LO 3NF are taken into account via a density-dependent
two-body potential ${\overline V}_{NNN}$ obtained by summing one
nucleon over the filled Fermi sea \cite{HKW09,HKW10}, and that is
added to the chiral N$^3$LO potential $V_{NN}$.  
The LECs $c_D$ and $c_E$ of the N$^2$LO 
chiral three-nucleon force are fixed so to reproduce to the binding
energies of $A=3$ nuclei and the $^3$H-$^3$He Gamow-Teller matrix element.  

The paper is organized as follows. 
In Sec.~2, we briefly describe the procedure we have followed to
choose the LECs of the $1 \pi$-exchange term $V_D$ and contact
interaction $V_E$, and the perturbative calculation of the properties
of infinite nuclear matter is outlined.
Our results and some concluding remarks are presented in Secs.~3 and 4,
respectively.
 
\section{Outline of calculations}
\subsection{Chiral potential}
In the last decade $NN$ potentials that are able to
reproduce accurately the $NN$ data have been derived in the framework of chiral
perturbation theory (ChPT) \cite{EM03,EGM05,ME11}.
In this work, we employ a chiral  N$^3$LO $NN$ potential with a cutoff
$\Lambda = 414$ MeV (published in Ref.~\cite{Cor07}), and using $n=10$
in the regulator function $f(p',p) = \exp [-(p'/\Lambda)^{2n}
- (p/\Lambda)^{2n}]$, i.e., a smooth, but rather steep cutoff function
is applied. 

In addition to this 2NF, we consider the contributions of a N$^2$LO
3NF. 
At this order the chiral 3NF is built up by three terms: a two-pion
exchange term, a one-pion exchange plus a 2N-contact interaction, and
a pure 3N-contact interaction. 
The two-pion exchange 3NF contains only LECs that are already present
in the $NN$ potential, while the last two terms
\begin{equation}
V_{D} = 
-\frac{c_D}{f^2_\pi\Lambda_\chi} \; \frac{g_A}{8f^2_\pi} 
\sum_{i \neq j \neq k}
\frac{\vec \sigma_j \cdot \vec q_j}{
 q^2_j + m^2_\pi }
( \mbox{\boldmath $\tau$}_i \cdot \mbox{\boldmath $\tau$}_j ) 
( \vec \sigma_i \cdot \vec q_j ) 
\label{eq_3nf_nnloc}
\end{equation}
and 
\begin{equation}
V_{E} = 
\frac{c_E}{f^4_\pi\Lambda_\chi}
\; \frac12
\sum_{j \neq k} 
 \mbox{\boldmath $\tau$}_j \cdot \mbox{\boldmath $\tau$}_k  \; .
\label{eq_3nf_nnlod}
\end{equation}
involve two new parameters $c_D$ and $c_E$, which do not appear in the
2N problem. 
There are many ways to fix these two parameters.
In the present work we have adopted a procedure that has been
recently introduced to constrain 
$c_D$ and $c_E$ \cite{Gardestig06,Gazit08,Marcucci12}. 
This procedure is based on the observation that the LEC $c_D$ appears
also in a two-nucleon contact
term in the $NN$ axial current operator derived in chiral EFT up to
N$^2$LO. Therefore, $c_D$ can be fixed to reproduce the accurate experimental
value of the triton $\beta$-decay half-life,
and in particular of its Gamow-Teller component (GT).
More precisely, we have first calculated the $^3$H and $^3$He
wave functions within the hyperspherical harmonics method (see Ref.~\cite{Kie08}
for a review), using the chiral 2NF plus 3NF presented above.
The LECs $c_D$ 
and $c_E$ are then determined by fitting the $A=3$ experimental binding
energies and the observed triton GT value. The values obtained are:
$c_D$ = -0.40 GeV$^{-1}$ and $c_E$ = -0.07 GeV$^{-1}$.

\subsection{Nuclear matter calculations}
We calculate the ground-state energy (g.s.e.) per nucleon of infinite
nuclear matter within the framework of many-body perturbation theory,
expressing the energy as a sum of Goldstone diagrams up to third order in
the interaction.

In order to take into account the effects of the N$^2$LO 3NF, a
density-dependent two-body potential ${\overline V}_{NNN}(k_F)$ is
obtained by summing one nucleon over the filled Fermi sea and added to
the chiral N$^3$LO potential $V_{NN}$. 
It is worth pointing out that, in the evaluation of the diagrams, the
matrix elements of ${\overline V}_{NNN}(k_F)$ have been multiplied by
a factor 1/3 in the first-order Hartree-Fock (HF) diagram, and by a factor 1/2 
in the calculation of the self-consistent single-particle energies.
This is done to take care of the correct combinatorial factors of the
normal-ordering at the two-body level of the 3NF \cite{HS10}.

\begin{figure}[H]
\begin{center}
\includegraphics[scale=0.6,angle=0]{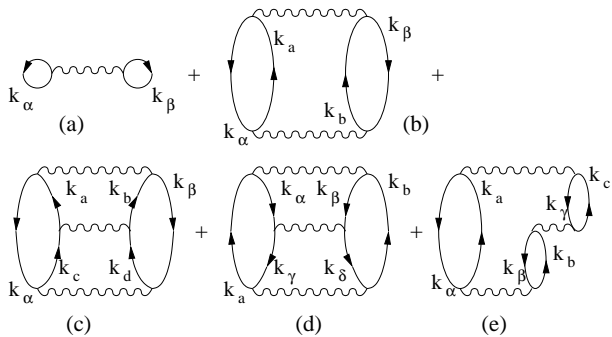}
\caption{First-, second-, and third-order diagrams of the Goldstone
  expansion included in our calculations.}
\label{figgold2+3}
\end{center}
\end{figure}

In Fig. \ref{figgold2+3} we report the diagrams we have included in our
calculation with $V_{NN}$ and ${\overline V}_{NNN}(k_F)$ vertices. The
contribution of the third-order particle-hole ($ph$) diagram (diagram
(e) in Fig. \ref{figgold2+3}), whose calculation is cumbersome, has
been included, at present, only for the SNM EOS.    

The analytic expressions of first-, second-, and third-order
particle-particle ($pp$) and hole-hole ($hh$) contributions may be
found in Ref. \cite{Coraggio13}, while the implicit expression of the
third-order $ph$ diagram is reported in Ref. \cite{MacKenzie69}.

Pad\'e approximants give an estimate of the value to which a
perturbative series may converge \cite{BG70}, therefore we have
calculated the $[2|1]$ Pad\'e approximant \cite{BG70} 

\begin{equation}
E_{[ 2|1 ]}=\mathcal{E}_0 + \mathcal{E}_1 +
\frac{\mathcal{E}_2}{1-\mathcal{E}_{3}/\mathcal{E}_2}~~,
\end{equation}

\noindent
$\mathcal{E}_i$ being the $i$th order energy contribution in the
perturbative expansion of the g.s.e., in order to study its convergence
properties.

\section{Results}
As mentioned in the previous Section, we calculate the energy per
particle of infinite nuclear matter in the framework of many-body
perturbation theory, including contributions up to third-order in
the interaction.
Therefore, it is worth studying the convergence of the
g.s.e. perturbative expansion. 

\begin{figure}[H]
\begin{center}
\includegraphics[scale=0.5,angle=0]{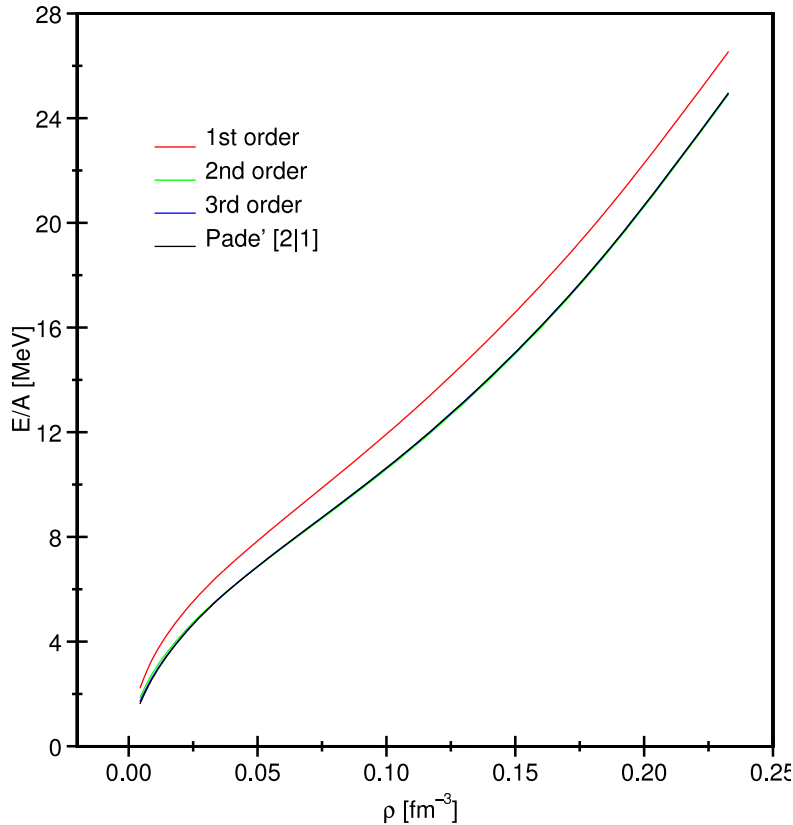}
\caption{(Color online) PNM energy per particle. The first,
  second, and third order in the perturbative expansion and the Pad\'e
  approximant $[2|1]$ are shown as a function of density.}
\label{convpnm}
\end{center}
\end{figure}

\begin{figure}[H]
\begin{center}
\includegraphics[scale=0.5,angle=0]{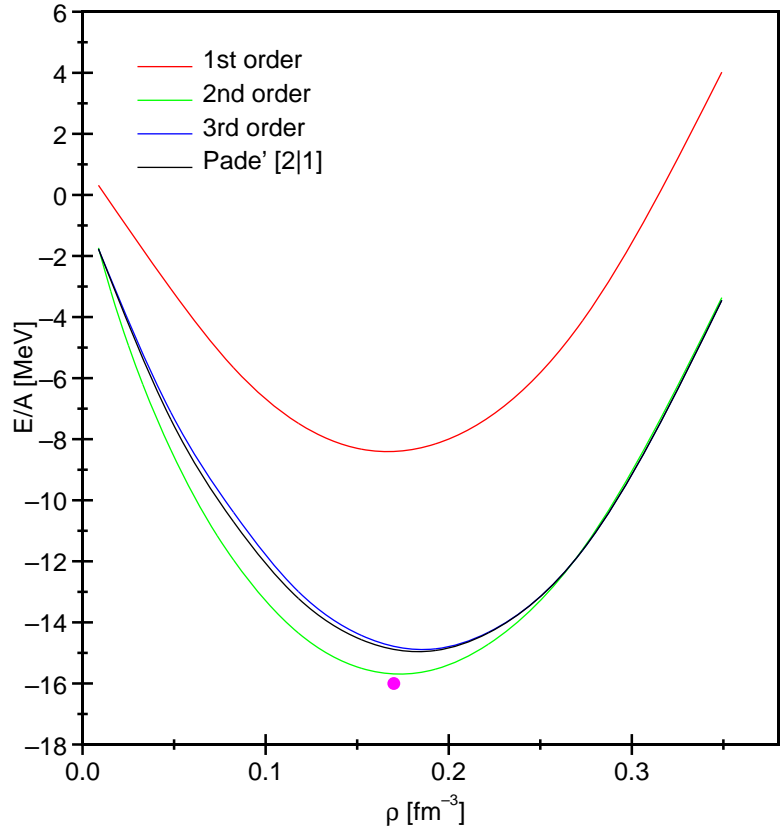}
\caption{(Color online) Same as in Fig. \ref{convpnm}, but for the SNM
  energy per particle.} 
\label{convsnm}
\end{center}
\end{figure}

In Fig.~\ref{convpnm} and \ref{convsnm} we show, respectively, the PNM
and SNM EOS as a function of density, calculated at various orders in
the perturbative expansion.
From Fig. \ref{convpnm} it can be seen that the PNM energy
per nucleon calculated at second order is almost indistinguishable
from the one computed at third order and consequently from the $[2|1]$
Pad\'e approximant for the whole range of density considered.    
As regards the SNM, we see from Fig. \ref{convsnm} that the EOS
calculated at second order does not differ much from the one
computed at third order, the latter being placed almost on the top of
the $[2|1]$ Pad\'e approximant.
This is a clear indication that the adopted chiral N$^3$LO
$NN$+N$^2$LO $NNN$ potential has a satisfactory perturbative behavior
for both PNM and SNM calculations.

\begin{figure}[h]
\begin{center}
\includegraphics[scale=0.5,angle=0]{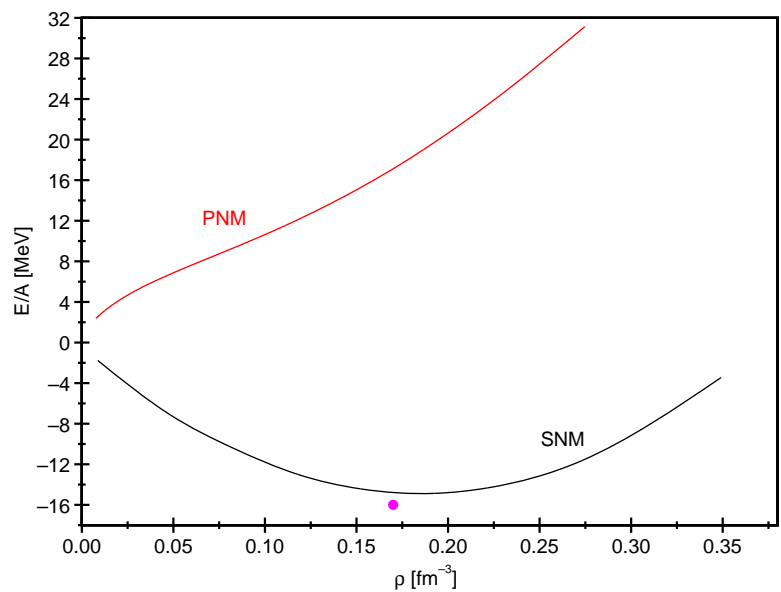}
\caption{(Color online) Results obtained for the PNM (red line) and
  the SNM (black line) energy per nucleon at third-order in
  perturbation theory.} 
\label{EOS}
\end{center}
\end{figure}

\begin{figure}[H]
\begin{center}
\includegraphics[scale=0.4,angle=0]{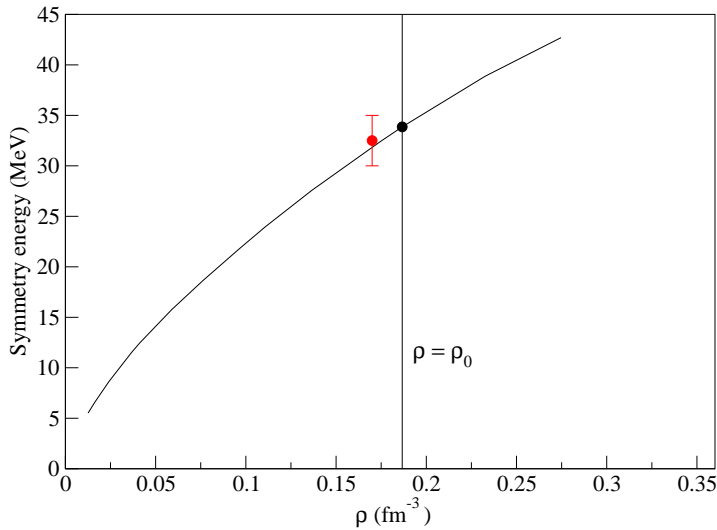}
\caption{(Color online) Calculated symmetry energy as a function of
  density. The red point represent the empirical value at saturation
  density \cite{Dutra12,Tsang12}.} 
\label{symen}
\end{center}
\end{figure}

The calculated PNM and SNM EOS and the corresponding symmetry energy
are shown in Fig. \ref{EOS} and \ref{symen}, respectively.
The inspection of these figures shows how our results reproduce well
the empirical SNM saturation point and the value of the symmetry
energy at saturation density, the latter quantity being related to the
isospin dependence of nuclear forces.
\vspace{1.0truecm}

Two other interesting physical quantities related to PNM and SNM at
saturation density are the incompressibility $K_0$  and the slope of
the symmetry energy $L$.

The empirical value of the SNM incompressibility is determined from
experimental data on Giant Monopole Resonance in finite nuclei, at
present its accepted value is $K_0 = 230 \pm 30$ MeV \cite{Dutra12}.
From our calculations we obtain a compressibility $K_0 = 279 $ MeV, a
value that is very close to the empirical one.

The quantity $L$ determines most of the behavior of the symmetry energy
in the proximity of saturation. 
Its empirical determination - centered at $L = 70$ MeV - is indirect
and mainly based on the analysis of heavy-ion collisions at
intermediate energies and nuclear structure measurements \cite{Tsang12}. 
Our calculated value for $L$ is 68 MeV, in good agreement with the
empirical value.

\section{Concluding remarks}
In this paper we have studied the properties of infinite nuclear
matter employing a chiral potential.
This has been done within the framework of the perturbative Goldstone
expansion, using a chiral N$^3$LO $NN$ and N$^2$LO $NNN$ potential
with a sharp cutoff $\Lambda=414$ MeV.
The LECs involved in the potential have been chosen consistently for
the two- and three-body components, and in particular the 3NF LECs
$c_D$ and $c_E$ have been fixed as to reproduce the experimental
$A=3$ binding energies and Gamow-Teller matrix element in triton $\beta$-decay.
Our results for PNM and SNM EOS turn out to be in good agreement with
the empirical properties of infinite nuclear matter. 
The ability to provide realistic nuclear matter predictions employing
(consistent) two- and three-body interactions whose LECs are constrained by the
properties of the two- and the three-nucleon systems is the main
outcome of our study, and this is certainly a very important point
that should be further investigated.

\section*{Acknowledgements}
This work was supported in part by the U.S. Department of Energy under
Grant No.\ DE-FG02-03ER41270 and No.\ DE-FG02-97ER-41014. 
We thank Norbert Kaiser for helpful discussions concerning the
calculation of the third-order particle-hole diagram.

\section*{References}
\bibliography{biblio}

\end{document}